# Quality Degradation Attack in Synthetic Data

Qinyi Liu[1][*], Dong Liu[2][**], Farhad Vadiee[1][* * *], Mohammad Khalil[1][†], and Pedro P. Vergara Barrios[2][‡]

[1]University of Bergen, Norway
[2]Delft University of Technology, Netherlands

**Abstract.** Synthetic Data Generation (SDG) can be used to facilitate privacy-preserving data sharing. However, most existing research focuses on privacy attacks where the adversary is the recipient of the released synthetic data and attempts to infer sensitive information from it. This study investigates quality degradation attacks initiated by adversaries who possess access to the real dataset or control over the generation process—such as the data owner, the synthetic data provider, or potential intruders. We formalize a corresponding threat model and empirically evaluate the effectiveness of targeted manipulations of real data (e.g., label flipping and feature-importance-based interventions) on the quality of generated synthetic data. The results show that even small perturbations can substantially reduce downstream predictive performance and increase statistical divergence, exposing vulnerabilities within SDG pipelines. This study highlights the need to integrate integrity verification and robustness mechanisms, alongside privacy protection, to ensure the reliability and trustworthiness of synthetic data sharing frameworks.

**Keywords:** synthetic data · secure and robust machine learning · Generative AI

## 1 Introduction

Synthetic data generation (SDG) is a machine learning process that learns the underlying distribution of real data to produce structurally and statistically similar synthetic datasets [7]. Synthetic data enables diverse applications, including privacy-preserving data sharing when real datasets cannot be disclosed, bias mitigation through data augmentation, and cost-effective annotation via synthetic labeled data [3] [17]. Its versatility and low cost have led to widespread adoption not only in healthcare, education, and autonomous driving, but also in many other domains, where it continues to yield promising results [3] [6].

---
[*] qinyi.liu@uib.no
[**] D.Liu-7@tudelft.nl
[* * *] farhad.vadiee@uib.no
[†] mohammad.khalil@uib.no
[‡] p.p.vergarabarrios@tudelft.nl

A critical challenge in using synthetic data as a privacy-enhancing technology for sharing sensitive datasets is ensuring that the generated data preserves privacy without leaking information under adversarial attacks. Prior studies have primarily explored privacy attacks, in which synthetic data receivers act as adversaries, employing techniques like membership inference attacks to reconstruct sensitive information from synthetic data [17] [4]. However, limited attention has been given to insider scenarios, where real dataset holders or synthetic data providers intentionally manipulate the data generation process to degrade synthetic data quality, reduce downstream task performance, and undermine its utility. This represents a significant gap in the synthetic data literature, as most existing work implicitly assumes that attacks are only conducted by receivers of synthetic datasets for privacy auditing purposes [16].

In this work, we address this gap by formalizing a threat model for quality degradation attacks in privacy-preserving synthetic data sharing scenarios. We consider adversaries with access to the real dataset—including the data owner (Client A) and the synthetic data provider (Provider B) or other parties that have access via system penetration. We evaluate the impact of targeted real-data manipulations on synthetic data fidelity and downstream predictive performance. Our contributions are threefold:

1. We formalize a practical threat model when synthetic data is used as privacy data sharing use case. This threat model captures attacker source, capabilities, and knowledge, extending the scope of synthetic data attacks beyond traditional privacy attacks.
2. We adapt existing quality degradation attack strategies, originally developed for general AI models, with minor modifications to fit the context of SDG. We then empirically assess their effectiveness across upstream, in-processing, and post-generation stages in synthetic data, showing that even modest perturbations can reduce downstream utility and increase statistical divergence.
3. We discuss broader implications for privacy-preserving data sharing and synthetic data research, highlighting the need for integrity verification and robustness mechanisms alongside conventional privacy protections.

The remainder of the paper reviews background (Sec. 2), defines the problem and threat model (Sec. 3), describes the experimental setup (Sec. 4), presents the results (Sec. 5), and concludes with future directions (Sec. 6).

## 2   Background and Related Work

This section first describes the technical background to understand quality degradation attack in synthetic data and related work by previous literature.

### 2.1   Synthetic data

As defined in the introduction, synthetic data is generated by a mathematical model or algorithm (i.e., a "generator") [5]. A Synthetic Data Generator is any algorithm (deterministic or stochastic) that takes an original dataset $D$ as input and outputs a synthetic dataset $S \sim \text{SDG}(D, m)$ ($D$ is the original dataset, $m$ is the number of samples to generate in the synthetic dataset) [16]. The types of synthetic data include text data, tabular data [1], time series data, multimedia data, such as images, audio and video. However, since this paper focuses on tabular datasets, the emphasis is placed exclusively on synthetic tabular data. In particular, we study using synthetic data in the privacy-preserving data sharing scenario, where synthetic data serves as a substitute for sensitive real datasets that cannot be disclosed directly.

**Synthetic data generation.** We evaluate three representative deep generative models for tabular data:

- **CTGAN** [7], a conditional tabular GAN that effectively handles mixed-type data and complex correlations;
- **TVAE** [18], a variational autoencoder tailored for tabular data with specialized loss functions for discrete/continuous columns;
- **NFLOW** [19] [2], a normalizing-flow model that offers exact likelihood computation and high expressiveness.

All models are used with their official implementations and default hyperparameters provided by the SDV [18] and SynthCity [19] libraries.

**Synthetic data evaluation.** We evaluate synthetic data quality from two perspectives that are standard in the literature:

- **Distributional fidelity**: Wasserstein Distance (WD), Kolmogorov–Smirnov statistic (KS), and Kullback–Leibler Divergence (KLD) between real and synthetic distributions.
- **Downstream machine-learning utility**: train-on-synthetic, test-on-real (TSTR) accuracy of three common classifiers (Logistic Regression, Random Forest, and MLP) on held-out real test data. Privacy evaluation is out of scope, as our threat model assumes the attacker is the data owner or provider who does not aim to leak information.

### 2.2 Related Attacks

**Quality-degradation attacks in standard ML** Data poisoning attacks that deliberately degrade model utility have been extensively studied in traditional machine learning [13]. Label flipping (LFA) and feature suppression/importance-based attacks (FIA) can reduce downstream accuracy by 20–50% with only 10–50% tampering ratio [12]. Prior work has not explored how classic poisoning primitives translate to synthetic data generation, and we provide the first systematic examination of this connection.

---

[1] Tabular data refers to data organized into tables, where information is arranged in rows and columns. seen from Krishnamurthi, S., Lerner, B.S. and Politz, J.G., 2017. Introduction to Tabular Data.

**Attacks on synthetic data generation** Virtually all prior attacks on tabular synthetic data target *privacy leakage* from the receiver's perspective, including membership inference [17],[9] attribute inference [16], reconstruction attacks [8] or re-identification attack[?]. In contrast, *quality degradation attacks initiated by the data owner or the generation provider* (i.e., insiders who control the real dataset or the SDG process) remain almost entirely unexplored. The very few existing works that touch upon quality degradation either (i) use synthetic data as an attack vector to poison downstream models [15], or (ii) study how poisoning of data downgrade the generative fidelity of Text-to-Image Diffusion Models [14]. This highlights a critical research gap: deliberate quality degradation attack that target the SDG itself remain overlooked in current literature.

## 3   Problem Setting & Threat model

### 3.1   Problem Setting

This paper focuses on the use of synthetic data as privacy-preserving data sharing, where real data cannot be directly shared, and synthetic data is employed as a substitute to enable access for stakeholders and external parties. As described in Section 2.1, this setting typically involves 2–3 parties: Client A, Provider B, and Receiver C. This process is illustrated in Figure 1.

- **Client A**: The owner of the real dataset. Due to privacy restrictions, Client A cannot directly share the real data, but instead generates or outsources the generation of synthetic data for sharing or exchange with related stakeholders (i.e. Receiver C).
- **Receiver C**: The end user(s) of the synthetic data. Receiver C may represent a single institution, multiple organizations, or even the administrators of an open data platform. Receiver C receive synthetic data either for publication or for use in downstream tasks. But due to privacy reasons, Client A cannot directly share real dataset with Receiver C.
- **Provider B (optional)**: A dedicated synthetic data generation company that may be involved when Client A outsources the task. As the SDG process grows more sophisticated and domain-specific, Provider B is expected to take on an increasingly important role.

In the privacy-preserving sharing scenario described, threats to synthetic data quality may originate at different points in the pipeline (e.g., from Client A's choice or manipulation of the real data source, or from a compromised or negligent Provider B). To study these threats rigorously we next formalize a threat model that captures attacker source (who tampers), capabilities (what manipulations are possible), and knowledge. Building on that formalization, this work addresses two core research questions:

- **RQ1 — Threat modeling.** How can realistic threat models for quality degradation be constructed for synthetic data use as privacy-preserving data sharing, specifically considering threats originating from the data owner (Client A) and the synthetic data provider (Provider B)?

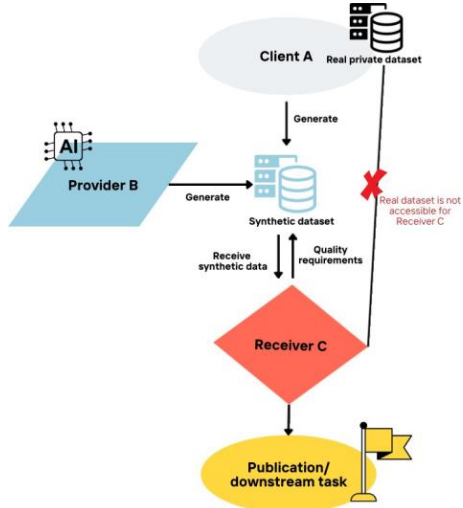

**Fig. 1.** Synthetic data outsourcing process

- **RQ2 — Effectiveness.** In the threat model of RQ1, to what extent can small or targeted manipulations of the real dataset induce measurable degradation in the synthetic data itself?

The following section (3.2 Threat Model) formalizes how threats from Client A vs. Provider B can be instantiated and operationalized to answer RQ1–RQ2.

### 3.2  Threat Model

This threat model directly addresses RQ1 introduced in Section 3.1. It formalizes the potential sources, capabilities, and constraints of attackers who aim to degrade the quality of synthetic data while respecting the privacy-preserving sharing framework.

**Attacker's Capability and Knowledge.** We assume the attacker can be either Client A (the holder of the real dataset), Provider B (a dedicated synthetic data generation entity, if involved), or a third party that has infiltrated their systems. The attacker has access to the real dataset $D$. This access may be:

- Direct, if the attacker is Client A.
- Legally granted for SDG and $D$, if the attacker is Provider B.
- Indirect, via system penetration, if the attacker is a third party.

The attacker can influence the generation of the synthetic dataset $\tilde{D}$ by modifying the input data (real dataset), SDG parameters, or post-processing outputs.

While the attacker does not know the exact verification mechanism used by Receiver C, he/she is aware that privacy constraints prevent direct access to $D$. The attacker cannot control downstream tasks of Receiver C. All actions are constrained mainly by computational resources, while detection is inherently difficult because Receiver C lacks access to the real dataset and therefore cannot reliably audit the synthetic data.

### 3.3  Attack Surfaces

The attacker's objective is to maximize the distributional divergence between synthetic ($p_{\tilde{D}}$) and real ($p_D$) data, thereby degrading the quality of the synthetic data:

$$\max_{\tilde{D}} D(p_{\tilde{D}}, p_D),$$

where $p_{\tilde{D}}$ and $p_D$ denote the synthetic and real data distributions, subject to a tampering ratio $|D_a|/|D| \leq r$, with $r \in \{10\%, 30\%, 50\%\}$.

The attacker's goal here is to maximize the distribution distance, rather than also including reduce the performance of downstream tasks. We set this way because the attacker cannot control the specific downstream tasks performed by Receiver C as stated before in attacker's capability. Furthermore, according to the literature, the greater the distribution gap between real and synthetic data, the worse the performance of synthetic data on downstream tasks [7]. Therefore, the attacker only needs to maximize the distribution gap to reduce the performance of synthetic data. Furthermore, the tampering ratio is set to account for the fact that in real-world scenarios, attackers may be limited by budgets.

We categorize quality-degradation attacks according to the stage of the SDG pipeline they target. Each attack surface is followed by concrete attack primitives (the taxonomy used in our experiments).

**Upstream attacks:** attacks that affect the real-data source or its selection before any generative training occurs. Examples include:
- **Incorrect or inappropriate data source:** choosing a mismatched cohort, stale snapshot, or otherwise wrong dataset that deviates from the intended population.
- **Real-data manipulation (poisoning):** LFA, targeted FIA, exemplar injection/removal prior to generator training.

**In-processing attacks:** attacks that influence the generation process itself (during SDG training or configuration). Examples include:
- **SDG parameter / resource manipulation:** forcing low training epochs, limiting model capacity, or configuring low-quality generator architectures (e.g., substituting GMM for a high-capacity model).

**Post-generation attacks:** attacks applied after synthetic data has been produced but before (or during) release. Examples include:
- **Distribution distortion:** altering summary statistics to cause large divergence (e.g., scale amplifications that dramatically increase KLD).

Specific implementation details of different statistical types and quantitative synthetic data quality assessment methods can be found in Section 4.

## 4 Experiments

In this section we provide the experimental details used to answer RQ2. Specifically, we describe the real-world datasets, the representative SDG models, the evaluation metrics for measuring distributional divergence and utility degradation. The implementation details of the attack primitives defined in Section 3.2.

**Datasets.** We used three tabular datasets that are widely used in machine learning and SDG research. Their basic information are summarized in Table 1.

**Table 1.** Description of datasets. The notations $C$, $B$, $M$ & $M_i$ represent number of continuous, binary, multi-class categorical, mixed variables and imbalance ratios (i.e., no. of minority samples/no. of majority samples) respectively.

| Dataset | Dataset size | Target Variable | $C$ | $B$ | $M$ | $M_i$ |
|---|---|---|---|---|---|---|
| Adult | 48k | "income>50K" | 8 | 1 | 4 | 0 |
| Law | 50k | "first_pf" | 8 | 2 | 4 | 0 |
| Loan | 21k | "Personal Loan" | 4 | 5 | 1 | 8 |

### 4.1 Attack implementation

We follow the attack-surface taxonomy introduced earlier and describe the exact manipulations, parameters and formalizations implemented in our experiments.

*Upstream attacks.* Inspired by data poisoning attacks, these attacks alter the real-data source or its selection prior to any generative training.

- **Incorrect / inappropriate data source.** As a baseline upstream degradation, we replace the intended real dataset with incorrect data sampled uniformly over a wide range and perturbed by small Gaussian noise. We generate incorrect data with uniform random values (range $[-10, 10]$) plus small Gaussian noise ($\sigma = 0.1$). This operation models the effect of a mismatched or erroneous source that deviates substantially from the target population used to define $p_D$.
- **Real-data manipulation (poisoning).** We implement two primary poisoning primitives applied to the training data $D_{\text{train}}$ prior to generator training:
  **LFA:** A fraction of labels indexed by $I_{\text{flip}}$ is inverted according to

$$D'[y][I_{\text{flip}}] = 1 - D_{\text{train}}[y][I_{\text{flip}}], \quad (1)$$

  where $D'$ denotes the poisoned dataset and binary labels are used for clarity.

**FIA:** We retain only a subset of critical features $F_{\text{selected}}$ (selected via a pre-trained Random Forest importance ranking) and discard or nullify other features. The poisoned dataset is thus

$$D^* = D_{\text{train}}\ F_{\text{selected}} \cup \{y\}, \quad (2)$$

which models an attack that removes or suppresses features deemed important for downstream utility. For consistency across attacks, we interpret the tampering ratio $r \in \{10\%, 30\%, 50\%\}$ in terms of the *effective fraction of data or features manipulated*. For LFA, $r$ directly denotes the percentage of training samples whose labels are flipped. For FIA, retaining the top $r\%$ of features is equivalent to suppressing the remaining $(1 - r)\%$; thus, smaller $r$ corresponds to stronger tampering in FIA (e.g., $r = 10\%$ reflects the most aggressive feature suppression), whereas larger $r$ corresponds to stronger tampering in LFA.

*In-processing attacks.* Inspired by adversarial attacks and model-degradation attacks, these attacks influence the SDG training process or the generator configuration.

- **Oversimplified SDG.** Using a Gaussian Mixture Model (GMM) with 10 mixture components to produce oversimplified data that deviates significantly from the true distribution.
- **Low-epochs (resource-constrained training).** We simulate constrained training by limiting the SDG training to a very small number of epochs (e.g., $E = 10$), instead of the 100 epochs used in other experiments. This setting produce undertrained generators whose outputs deviate from $p_D$.

*Post-generation attacks.* Inspired by adversarial attack but move it to the generation output, these attacks are applied after synthetic data have been produced but before release.

- **Distribution distortion (output scaling).** As a simple and effective post-release tampering, we scale numeric features in the synthetic dataset by a constant factor (e.g., factor $s = 2$), which distorts summary statistics and increases divergence measures such as KLD:

$$\tilde{x}_{\text{scaled}} = s \cdot \tilde{x}, \quad s = 2.$$

- **Noise injection to synthetic data.** We also generate directly degraded synthetic outputs by injecting large Gaussian noise into generated values:

$$\tilde{x} = x_{\text{gen}} + \mathcal{N}(0, \sigma^2), \quad \sigma = 10,$$

where $x_{\text{gen}}$ is an originally generated value. This primitive captures cases where an attacker intentionally perturbs the generator outputs during training.

## 5  Results

In the experiments below (Table 2), we observe that real-data manipulation (poisoning) methods such as FIA and LFA can significantly degrade the quality of synthetic data. All three downstream predictive metrics (LR, RF and MLP) show negative changes, indicating that poisoning of the real dataset reduces downstream predictive performance when models are trained on poisoned synthetic data compared to clean baselines. At the same time, the three similarity/divergence metrics (WD, KS and KLD) increase substantially, reflecting a larger dissimilarity between the released synthetic distribution and the intended real distribution. Specifically, compared with clean datasets, FIA increases WD by 1,337.55% while reducing MLP accuracy by 11.57%, whereas LFA increases WD by 322.34% and reduces MLP accuracy by 17.08% (Table 2). These results corroborate prior findings that synthetic-data quality is tied to both statistical fidelity and predictive accuracy [1], and they demonstrate that poisoning attacks operate by maximizing distributional shifts that in turn harm predictive utility.

The impact of the attack ratio $r$ differs between the two attacks. For LFA, higher tampering ratios correspond to stronger attacks and greater performance degradation: for instance, increasing $r$ from 10% to 50% leads to progressively larger drops in downstream model accuracy (MLP decreases from -2.87% to -44.03%) and substantial increases in distributional metrics. In contrast, for FIA, lower ratios are more aggressive because $r$ represents the fraction of top features *retained*; thus, 10% represents the most extreme feature suppression, producing the largest increases in WD (2926.74%) and decreases in downstream accuracy (MLP -8.89%). Overall, these results show that the severity of the poisoning attack is closely tied to the tampering ratio, with its effect being proportional to $r$ in LFA and inversely proportional in FIA, demonstrating that both the amount of label flipping and the degree of feature suppression critically determine synthetic data quality and downstream predictive performance.

Table 3 reports averaged effects of a broader set of quality-degradation attacks. The impact varies substantially by attack primitive. Simple generator degradation ("Oversimplified SDG") and resource constraints ("Low epochs") induce only modest changes in divergence and small utility drops (e.g., Oversimplified SDG: LR $-2.10$%, MLP $-1.36$%, WD 7.06), suggesting these modes are less destructive but also harder to detect via large divergence signals. By contrast, source-level and post-generation manipulations produce dramatic divergence and utility loss: replacing the real source with an inappropriate dataset yields large WD/KLD increases (WD = 1335.48, KLD = 5747.71) and severe MLP degradation ($-55.53$%), while noise injection and distribution-distortion also produce very large divergence values (e.g., noise: WD = 1079.37, scaling: WD = 856.31) with heterogeneous effects on different classifiers (e.g., noise causes a large LR drop of $-41.33$% but smaller changes for RF/MLP). The averaged row summarizes these behaviors: substantial mean divergence increases (WD = 475.36, KLD = 2411.90) accompany notable average reductions in classifier performance (LR = $-17.30$%, RF = $-2.68$%, MLP = $-12.86$%).

**Table 2.** Average Percentage Changes of General Synthetic Data Generators under FIA and LFA. Values are averaged over three SDGs for three datasets.

| Attack Type | Attack Ratio | LR | RF | MLP | WD | KS | KLD |
|---|---|---|---|---|---|---|---|
| FIA | 0.1 | -2.03 | -11.75 | -8.89 | 2926.74 | 984.52 | 3584.47 |
|  | 0.3 | -2.04 | -6.38 | -2.77 | 649.95 | 711.79 | 4274.13 |
|  | 0.5 | -24.62 | -21.51 | -23.04 | 430.21 | 532.79 | 2578.80 |
|  | Avg | -9.56 | -13.21 | -11.57 | 1337.55 | 743.70 | 3479.04 |
| LFA | 0.1 | -1.32 | -4.69 | -2.87 | 388.21 | 405.47 | 2433.61 |
|  | 0.3 | -2.05 | -9.38 | -4.34 | 283.95 | 395.51 | 2570.84 |
|  | 0.5 | -43.18 | -41.04 | -44.03 | 294.86 | 375.84 | 2371.45 |
|  | Avg | -15.52 | -18.37 | -17.08 | 322.34 | 392.27 | 2458.63 |

*Note.* Negative values in LR, RF, and MLP indicate performance drops in downstream predictive models. While larger values of WD, KS, and KLD denote greater dissimilarity between synthetic and real datasets indicate performance drop.

**Table 3.** Average percentage changes under different quality degradation attacks. Values are averaged over three SDGs and three datasets. Negative values denote performance drops.

| Attack Type | LR | RF | MLP | WD | KS | KLD |
|---|---|---|---|---|---|---|
| Oversimplified SDG | -2.10 | -1.21 | -1.36 | 7.06 | 277.33 | 404.84 |
| Low epochs | -1.12 | -1.59 | -1.61 | 99.60 | 119.11 | 70.80 |
| Incorrect / inappropriate data source | -23.79 | -8.59 | -55.53 | 1335.48 | 675.55 | 5747.71 |
| Noise injection | -41.33 | -0.82 | -2.17 | 1079.37 | 394.58 | 1244.08 |
| Distribution distortion | -18.17 | -1.17 | -1.64 | 856.31 | 418.95 | 4588.07 |
| **Average** | -17.30 | -2.68 | -12.86 | 475.36 | 377.90 | 2411.90 |

*Note.* Negative values in LR, RF, and MLP indicate performance drops in downstream predictive models. While larger values of WD, KS, and KLD denote greater dissimilarity between synthetic and real datasets indicate performance drop.

## 6  Discussion and Conclusion

This work examined quality-degradation attacks on SDG pipelines in privacy-preserving data sharing. Regarding **RQ1**, our empirical analysis validates the proposed threat model, showing that realistic degradation scenarios must account for two distinct upstream adversaries: the *data owner* (Client A) and the *synthetic data provider* (Provider B), both of whom may influence real data or training processes. Addressing **RQ2**, our experiments demonstrate that even small or targeted manipulations to the real dataset can induce substantial drops in synthetic data fidelity and downstream predictive utility. Subtle label or feature perturbations propagate through the generative process, leading to pronounced distributional divergence and performance degradation across models. However, divergence does not uniformly predict utility loss, reflecting heterogeneous classifier sensitivity and highlighting the need to consider both statistical and predictive metrics when evaluating poisoning effects.

   **Limitations.** Our attacks are intentionally simple and targeted at tabular data. Extending the framework to other modalities (e.g., image backdoors or adversarial perturbations) remains an important direction for broader applicability.

**Implications for privacy-preserving data sharing.** Our findings reveal a critical challenge for receivers of synthetic data in privacy-preserving settings. Since Receiver C does not have access to the underlying real dataset, conventional evaluation approaches—whether privacy metrics or statistical similarity measures—cannot rely on direct comparisons between synthetic and real data [11]. As a result, assessing the quality of synthetic data solely based on the released dataset may be misleading, because any degradation that has already occurred could remain undetected. This underscores the need for integrity verification strategies that do not require full access to the real data, such as provenance tracking, auditing SDG configurations, or post-hoc consistency checks, to ensure that synthetic datasets remain both reliable and privacy-preserving.

**Implications for the broader literature and future work.** From a broader research perspective, this study extends the synthetic data security literature beyond its traditional privacy-centric focus. Prior research (e.g. [16] [17]) has overwhelmingly treated attacks as diagnostic tools for auditing privacy leakage implicitly positioning the adversary as the receiver of synthetic data. By contrast, our work expands the threat landscape to include upstream and in-processing adversaries, who possess or influence the real dataset itself. The experimental results of our RQ2 show that slight changes may lead to lower quality of synthetic data, which also proves that it is valuable to study this new threat model at the practical level. This reframing of threat model introduces a complementary dimension to SDG risk analysis: rather than only measuring how much private information leaks *from* synthetic data, we can ask how much utility and fidelity can be degraded *within* it.